\documentclass[twocolumn,english,aps,superscriptaddress, pra]{revtex4-1}

\usepackage[latin9]{inputenc}
\usepackage{amsmath}
\usepackage{amssymb}
\usepackage{graphicx}
\usepackage{babel}
\usepackage{mathrsfs}
\usepackage{amsfonts}

\numberwithin{equation}{section}
\renewcommand\theequation{\arabic{section}.\arabic{equation}}

\begin{document}

\title{Quantum N-Boson States and Quantized Motion of Solitonic Droplets: \\ Universal Scaling Properties in Low Dimensions}

\author{ Jeff Maki} 
\affiliation{Department of Physics and Astronomy, University of British Columbia, Vancouver V6T 1Z1, Canada}

\author{Mohammadreza Mohammadi}
\affiliation{Department of Physics, 60 St. George St., Toronto, Ontario M5S 1A7, Canada}

\author{Fei Zhou} 
\affiliation{Department of Physics and Astronomy, University of British Columbia, Vancouver V6T 1Z1, Canada}

\date{Jan 02, 2014}

\begin{abstract}

In this article, we illustrate the scaling properties of a family of solutions for N attractive bosonic atoms in the limit of large N. These solutions represent the quantized dynamics of solitonic degrees of freedom in atomic droplets. In dimensions lower than two, or $d=2-\epsilon$, we demonstrate that the number of isotropic droplet states scales as $N^{3/2}/\epsilon^{1/2}$, and for $\epsilon=0$, or $d=2$, scales as ${N^2}$. The ground state energies scale as $N^{2 / \epsilon + 1}$ in $d=2-\epsilon$, and when $d=2$, scale as an exponential function of N. We obtain the universal energy spectra and the generalized Tjon relation; their scaling properties are uniquely determined by the asymptotic freedom of quantum bosonic fields at short distances, a distinct feature in low dimensions. We also investigate the effect of quantum loop corrections that arise from various virtual processes and show that the resultant lifetime for a wide range of excited states scales as $N^{\epsilon/2}E^{1-\epsilon/2}$. 

\end{abstract}

\maketitle

\section{Introduction}

Few-body correlations are known to play a fascinating role in a variety of quantum resonance phenomena~\cite{Skorniakov57,Faddeev61,Weinberg64,Yakubovsky67,Efimov70,Tjon75,Weinberg90,Kraemer06,Zaccanti09,Pollack09,Esry99,Chin10}. Since the discovery of Efimov states in the seventies \cite{Efimov70}, theoretical efforts have drastically increased to further our understanding of few-body physics. One of the important theoretical developments is perhaps the effective field theory approach to few-body problems~\cite{Kaplan98,Bedaque99,Mohr06,Nikolic07,Weinberg05} which relates resonant scattering phenomena to scale invariant critical points~\cite{Wilson71}. And thanks to the experimental observations of Feshbach resonances in cold gases, our interests in few-body states have been substantially revived during the last decade. Impressive experimental developments of inelastic loss spectroscopy near resonance have made it possible to perform precision measurements of loss rates. These breakthroughs have been successfully utilized to detect Efimov trimers \cite{Kraemer06,Zaccanti09,Pollack09,Esry99,Chin10}.

One can then ask the question what is beyond trimers, or the $N=3$ case, in the general context of quantum N-boson problems. Partially stimulated by the Feshbach resonance experiments, efforts have been made to understand few-body physics either beyond the Efimov paradigm or few-body clusters associated with Efimov states. The discovery and observation of associated universal four-body states, known as tetramers, appear to be another successful story in few-body research~\cite{Hammer07,Stecher08,Ferlaino09,Schmidt10}. However, going beyond four-body states so far appears to be extremely challenging and, not surprisingly, little is known about N-boson states with $N \gg 4$~\cite{Stecher09, Hanna06, Kievsky27}. On the other hand, from the point of view of cold gases, it is essential to understand the underlying few-body clusters as they are the fundamental building blocks for many-body correlations.

One of the main hurdles in the three-dimensional N-boson problem, which still remains to be an almost uncharted territory, is that in N-boson bound states as atoms get closer together, the resonant attractive interactions become more and more dominating.  This is evident if one simply considers particles interacting with a {\em weak} attractive contact interaction of strength $g_2$. The dimensionless two-body coupling constant $\tilde{g} (L)$ which is a measure of the ratio between $U(L)=g_2/L^d $, the two-body interaction energy at scale L, and $T(L)=1/2 L^2$, the kinetic energy at the same scale, has the following scaling property,

\begin{equation}
\tilde{g}(L) \sim \frac{g_2}{L^{d-2}},
\label{scaling}
\end{equation}

\noindent where $d$ is the spatial dimension. Although strictly speaking Eq.~(\ref{scaling}) is only valid when $\tilde{g}$ is much less than unity, nevertheless it is clear that at shorter scales the interaction becomes more dominating in $d > 2$~\cite{Tan08}. What further complicates the N-boson problem in 3D is that at the strong coupling fixed point of $g_2$, or near resonance, the three-body interactions also exhibit discrete scale invariance due to the renormalization flow \cite{Bedaque99, Mohr06}. This peculiar feature implies an additional dependence of the N-boson physics on a non-universal ultraviolet regime that is consistent with numerical findings in Ref.~\cite{Stecher09, Hanna06, Kievsky27} \cite{Jiang13}. For $d < 2$, the opposite occurs; namely the system becomes free when approaching smaller and smaller scales, that is it becomes {\em asymptotically free}, similar to quantum chromodynamics~\cite{Gross73,Poltizer73}.

The general scale dependence of $\tilde{g}$ can be systematically obtained by analysing the standard renormalization group equations. We restrict ourselves to a positive effective scattering length or bound state size $a$ due to an attractive contact interaction in low dimensions. For a given $a$, the corresponding interaction strength in $d=2-\epsilon$ dimension is,

\begin{equation}
g_2 =- \frac{1}{a^\epsilon}\frac{(4\pi)^{1-\epsilon/2}}{\Gamma(\epsilon/2)},
\label{g2}
\end{equation}

\noindent where $\Gamma$ is the Gamma function and diverges as $1/\epsilon$ near $\epsilon=0$. Following a similar calculation to those in Ref. \cite{Zhou13}, one finds for $d=2-\epsilon$,

\begin{equation}
\tilde{g}(L) = -(\frac{L}{a})^{\epsilon}\frac{(4\pi)^{1-\epsilon/2}}{\Gamma(\epsilon/2)} 
(1 - (\frac{L}{a})^{\epsilon}\frac{ \sin\pi \epsilon/2}{\pi \epsilon/2})^{-1}.
\end{equation}

In 3D, $\tilde{g}(L)=\frac{4\pi a}{L}(1-\frac{2a}{\pi L})^{-1}$. Indeed, $\tilde{g}(L\rightarrow\infty)$ vanishes at large distances, consistent with the naive scaling argument presented above. However, the behaviour at distances much shorter than $a$ is set by the asymptotic value $\tilde{g}(L\rightarrow 0)=-2\pi^2$ indicating the relevance of interactions. In (2-$\epsilon$)D, the corresponding limit yields $\tilde{g}(L\rightarrow 0)=- (L/a)^{\epsilon}$ illustrating the asymptotic freedom of bosons. Moreover, in 1D, the value of $\tilde{g}(L)$ in the infrared limit approaches a constant value which implies a strongly interacting regime in the dilute limit.

For N-boson systems, the total interaction energy is $N^2 U(L)$, while the kinetic energy  scales as $N T(L)$.  At short distances, the ratio between these two energies thus scales as $N \tilde{g}(L \sim 0)$. In 3D, the magnitude of this quantity becomes much larger than unity as both $N$ is large and $\tilde{g}(L\rightarrow 0)$ is a constant of order unity. Thus in 3D the N-boson states are dictated by the mutual attractions and the total attractive energy, $ N^2 \tilde{g}(L) T(L)$ where $T(L)$ scales as $1/L^2$, becomes more and more negative as one enters the ultraviolet limit. Strictly speaking for N bosons interacting via a contact interaction, its energy is not bounded from below because of this ultraviolet catastrophe ({\it UC}). To cure the {\it UC}, one has to further regularize the theory by specifying the details of the potential at short distances. One anticipates that its properties are not universal, and very much depend on how bosons interact at these very short distances. For this reason, quantum scalar fields with attractive contact interactions are usually considered to be {\it sick} theories. This aspect of the problem and the issue of {\it UC} in 3D are not totally surprising. In fact it has been known for the $N=3$ case; namely the energy scale that sets the binding energies of the Efimov trimers is given by a non-universal three-body parameter. 

Fortunately, the catastrophe mentioned above does not appear in dimensions lower than two. Although what happens in 3D at the $N$-body level ($N \gg 4$) still remains to be explored and understood, in this article we report the results on universal N-boson droplet states in 2-$\epsilon$ dimensions ($\epsilon >0$). There are two essential ingredients in our analysis of the scaling properties of N-boson states. First of all, particles become asymptotically free or non-interacting at short distances in low dimensions. This is in contrast to 3D where near-resonance attractive interactions are asymptotically dominating in the ultraviolet limit. Thus, in low dimensions at the ultraviolet scales the energetics is predominately determined by the kinetic energy associated with the uncertainty principle. Indeed, $\tilde{g}(L) \sim (\frac{L}{a})^\epsilon$ in low dimensions and vanishes as L goes to zero. For any large but fixed value of $N$,  $N \tilde{g}(L)$, the ratio between the total interaction energy relative to the kinetic energy, becomes vanishingly small as $L$ approaches zero. So unlike in 3D, here the total energy, which now is mainly the kinetic energy, $N T(L)$, {\it increases} as $L$ approaches zero. As a result, this system of N bosons interacting via a contact interaction can be shown to be bounded from below (see more discussions below). Physically, it is the dominance of the kinetic energy at short distances that effectively keeps particles from collapsing into a non-universal ultraviolet regime; it removes the {\it UC} in low dimensions. The universal scaling property obtained below is a manifestation of the uncertainty principle in quantum mechanics. Secondly, as a consequence of asymptotic freedom, via applying the standard effective potential method, one can further show that the induced or renormalized three-, four-body interactions etc. also become irrelevant at short distances. Technically, it is this property that makes it possible for us to derive a universal effective potential for N-boson states and carry out the scaling analysis.

In 2D, the limit of $N=3,4$, i.e. three- and four-body bound states have been studied previously~\cite{Bruch79,Nielsen99,Hammer04,Platter04,Blume05,Brodsky06,Nishida08}. The authors of Ref.~\cite{Hammer04} have also proposed a bound state for N bosons; we later identify it as a classical solution to the 2D quantum problem studied below. We will return to discuss the relation between our solutions and these previous results at a later point.

Our analysis of the N-boson problem is organized as follows: In Section II, the field theoretic framework for a Bosonic field is put forth, and an ansatz for studying the solitonic motion of the system is discussed. In Section III the solitonic motion is quantized and the scaling properties of these solitonic N-boson states are found. Sections IV and V examine the effect of loop corrections to the quantized solitonic N-boson states. These results are then compared to the variational energy of condensates at different length scales in Section VI. Finally, in Section  VII we summarize our results and discuss the relevance to cold atoms.

\section{Effective Theory for the Scaling Variable $\lambda$}

In order to study the scaling properties of bound states, it is most convenient to formulate the problem in terms of functional integrals. The Lagrangian for a bosonic field $\psi$ is given below,

\begin{equation}
\mathcal{L}=\int d{\bf r} \big( i \psi^* \partial_t \psi - \psi^* (-\frac{\nabla^2}{2})\psi -\frac{g_2}{2} \psi^*\psi^* \psi\psi \big),
\end{equation}

\noindent where $g_2$ is the two-body interaction constant defined in Eq.(\ref{g2}) and $m=\hbar=1$. For the purpose of studying few-body physics, we choose to work with the density ($\rho$) and phase ($\phi$) fields and
$\psi=\sqrt{\rho} \exp(i\phi)$. The corresponding Lagrangian is 

\begin{equation}
\mathcal{L}=-\int d{\bf r} \big(\rho \partial_t \phi +\frac{1}{2}\nabla \sqrt{\rho} \cdot \nabla \sqrt{\rho}  +\frac{\rho}{2}\nabla \phi \cdot \nabla \phi  +\frac{g_2}{2} \rho\rho \big).
\nonumber
\end{equation}

\noindent This Lagrangian leads to the following semiclassical equations of motion

\begin{eqnarray}
\partial_t \phi-\frac{1}{2\sqrt{\rho}}\nabla^2 \sqrt{\rho}+\frac{1}{2}\nabla\phi\cdot \nabla \phi + g_2\rho=0, \nonumber \\
\partial_t \rho +\nabla \cdot (\rho \nabla \phi)=0.
\end{eqnarray}

We then expand the fields around a semiclassical {\it isotropic} solution to the equation of motion, $\rho_{sc},\phi_{sc}$ as $\rho=\rho_{sc}+\rho_{qc}$ and $\phi=\phi_{sc}+\phi_{qc}$. The subscripts {\em sc} and {\em qc} indicate semiclassical and quantum correction fields respectively. For isotropic solitonic solutions, $\rho_{sc}$ and $\phi_{sc}$ should have the following scaling form

\begin{eqnarray}
\rho_{sc}(r, t)& = &\frac{N}{\lambda^d(t)}f(\frac{r}{\lambda(t)}), \nonumber \\
\phi_{sc}(r,t) & = &\frac{r^2}{2}\frac{\dot\lambda}{\lambda} +\int dt \frac{B_1}{2\lambda^2(t)} -\frac{N B_2}{\lambda^d},
\label{f-function}
\end{eqnarray}

\noindent where $r$ is the distance measured from the center of the mass of the system. The $f$-function satisfies the normalization condition, $\frac{2\pi^2}{\Gamma(d/2)}\int dx x^{d-1} f(x) =1$, and $B_1=\frac{1}{4f(0)} \partial^2_x f({x})|_{x=0}$ and $ B_2=g_2 f(0)$ are two constants that depend on the properties of the scaling function $f(x=r/\lambda)$  at $r=0$. A similar scaling ansatz has been used in previous works \citep{Timmermans00, Zoller97, Sackett98} to study the dynamical properties, such as collapse times, of condensates in higher dimensions and in the presence of trapping potentials. As mentioned above, this scaling ansatz emphasizes the solitonic nature of the system. The ansatz in Eq.~(\ref{f-function}) is a generalization of the textbook 1D soliton solution~\cite{Lakshmanan03, Zakharov72}. Thus the resulting N-boson droplet states are actually N-boson solitonic droplet states (or simply N-boson states for brevity).

The partition function for this system can then be written as

\begin{eqnarray}
Z &\approx &\int D\lambda \int D\rho_{qc} D\phi_{qc} \nonumber \\
& & \exp(i \int dt \mathcal{L}_{sc}(\lambda(t))+ \mathcal{L}_{qc}(\{\rho_{qc}\},\{\phi_{qc}\})). \nonumber\\
\end{eqnarray}

\noindent The explicit forms of $\mathcal{L}_{sc,qc}$ are given in Appendix A. After integrating out the fast fluctuation fields $\rho_{qc}$, $\phi_{qc}$, we then obtain an effective Lagrangian for the dynamical scaling variable $\lambda(t)$ in $d=2-\epsilon$ dimensions,

\begin{eqnarray}
Z& \approx &\int D\lambda \exp(i\int dt \mathcal{L}_{eff}), \nonumber \\
\mathcal{L}_{eff} &=& \mathcal{L}_{sc}+\mathcal{L}_{qc},\nonumber \\
\mathcal{L}_{sc} &=& C_1 \frac{N}{2} (\partial_t {\lambda})^2 -C_2 \frac{N}{2\lambda^2}+C_3 \epsilon \frac{N^2}{\lambda^{2-\epsilon}a^\epsilon}, \nonumber\\
\mathcal{L}_{qc} & =& \frac{1}{2}\sum_{k > 2\pi/\lambda} (\epsilon^2_k + 2 \epsilon_k \mu)^{1/2}-({\epsilon_k+\mu}),
\label{qc}
\end{eqnarray}

\noindent where  $\mu = N/\lambda^{d} / {g}_2$. $C_{1,2,3}$, as well as $C_4$, defined below, are dimensionless constants further depending on the details of the scaling function $f(r/\lambda)$ introduced in Eq.~(\ref{f-function}). Although the spectrum is completely known only when $C_{1,2,3,4}$ are available, most of the scaling properties discussed below are robust and depend little on the specific values of these coefficients. In Appendix A, one can find values of $C_{1,2,3}$ calculated for different ansatzs of the scaling function $f(x) $.

The explicit form of $\mathcal{L}_{qc}$ in Eq.~(\ref{qc}), was determined under the assumption that the density profile was constant up to a size $\lambda$. This allowed the fluctuations to be expanded in terms of plane waves with wave vectors that have magnitudes, $k$, larger than $2 \pi / \lambda$. If another choice of boundary conditions is used, the only significant modification to the fluctuation spectrum will be to the modes that have wave vectors with magnitudes $k \approx 2 \pi / \lambda$. However as will be seen below, the dominant contributions to $\mathcal{L}_{qc}$, are from modes with $k \sim \sqrt{\mu}$, where the phonon modes start to resemble the spectrum of a free particle and form a continuum. As a result, this choice of boundary conditions does not alter the following analysis. Further discussions of this approximation are found in Appendix B.

Two remarks are now in order. Since we are dealing with attractive interactions, the chemical potential and compressibility are both negative. Thus the energy of phonons (or $\mathcal{L}_{qc}$) always carries an imaginary part when $|\mu| \gg 1/\lambda^2$ or $\epsilon N(\lambda/a)^\epsilon \gg 1$. This reflects an instability of density waves at an {\em arbitrary} scale larger than $\lambda_0={a}(\epsilon/{N})^{1/\epsilon}$. Only if ${\lambda}$ is less than  $\lambda_0$ do phonons become fully stable. Not surprisingly, $\lambda_0$ is precisely the length associated with the classical equilibrium position $\lambda_e$.

Equally as important, the leading quantum correction  $\mathcal{L}_{qc}$ in Eq.(\ref{qc}) precisely describes the contributions of zero point fluctuations of {\em phonons} at scales smaller than $\lambda$ (as indicated by the lower cut-off in Eq.~(\ref{qc})). Since only the fluctuations with wavelengths shorter than the scaling parameter $\lambda$ contribute, splitting the density and phase fields into semiclassical and fluctuating parts is equivalent to separating fast fluctuating fields from the slow motion at scale $\lambda$. 

Following $\mathcal{L}_{sc}$, the interaction energy at a scale of $\lambda$ is given as $C_3 N^2/\lambda^d \ {g}_2$ while the kinetic energy associated with $\lambda$ is $C_2 N/2  \ (\partial_t\lambda)^2$. The velocity $\partial_t \lambda$  from the conversion of the interaction energy above is therefore proportional to $\sqrt{N {g}_2}/\lambda^{d/2}$. The duration of the motion at scale $\lambda$ is thus

\begin{equation}
\tau_0(\lambda)=\lambda^{d/2+1}\frac{1}{\sqrt{N g_2}}.  
\label{duration}
\end{equation}

\noindent This turns out also to be the time scale associated with the fluctuating fields at scale $\lambda$, the slowest fluctuation modes. Indeed, the fluctuations at wave vector $k$  have a corresponding frequency $\sqrt{\mu}k$ or $\sqrt{N g_2}\lambda^{-d/2}k$. For attractive interactions, negative $g_2$, the frequency is naturally imaginary, which indicates the time scale for an instability to occur at that particular wave vector. If $2 \pi / k$ is chosen to be $\lambda$, the longest wavelength for microscopic fluctuations for a given scaling parameter (which is also equivalent to $\lambda$), then the time scale associated with the onset of instability is equivalent to the duration of motion at scale $\lambda$ defined in Eq.~(\ref{duration}). 

By restricting ourselves to the fluctuations at scales shorter than $\lambda$, we effectively separate the fast motions of fluctuation fields $\rho_{qc}, \phi_{qc}$ from the slow motion of the semiclassical fields $\rho_{sc}$, $\phi_{sc}$ at the the scaling parameter $\lambda$. In other words, we only need to take into account the quantum correction due to fields that fluctuate faster than the time scale $\tau_0(\lambda)$.

The N-boson quantum dynamics is described by this effective Lagrangian. Quantization of the solitonic motion associated with $\lambda$ can be carried out using the standard Feynman path integral method and below we show the result of quantization.

\section{Quantized Solitonic Droplet Dynamics and N-boson States}

\subsection{$D=2-\epsilon$}

The partition function naturally describes the Feynman path integral of the dynamics of $\lambda(t)$. We quantize the Hamiltonian by applying the standard canonical quantization technique:

\begin{equation}
\left[ \lambda, P_{\lambda} \right] = i \hbar.
\label{canonical_quantization}
\end{equation}

\noindent The corresponding quantum Hamiltonian for $\lambda$ is

\begin{eqnarray}
\mathcal{H} &=& \frac{1}{2C_1 N} P^2_\lambda+ C_2 \frac{N}{2\lambda^2}-C_3 \epsilon \frac{N^2}{\lambda^{2-\epsilon}a^\epsilon}+\delta \mathcal{H} \nonumber \\ 
\delta \mathcal{H} &=& C_4 \frac{(\epsilon N)^{2-\epsilon/2}}{2 a^2} (\frac{a}{\lambda})^{(2-\epsilon)^2/2} F(\frac{2^{1-\epsilon}}{\pi^{1+\epsilon/2}} \frac{1}{\Gamma(\epsilon)} 
N (\frac{\lambda}{a})^{\epsilon}). \nonumber\\
\label{Hamiltonian}
\end{eqnarray}

\noindent And

\begin{equation}
F(\eta^2)=\frac{2\pi^d}{\Gamma(d/2)} \int_{1/\eta}^\infty dx x^{d-1} [\sqrt{x^4-2x^2}-x^2+1],
\label{F}
\end{equation}

\noindent which is a convergent integral for $d < 2$ and approaches a constant of order unity as $\eta$, $N\tilde{g}$, approaches infinity. Since $F(\eta)$, as shown in Appendix B, is imaginary for a wide range of $\lambda$ values, it implies an instability associated with the corresponding scale $\lambda$. Naturally, it yields an estimate of the lifetime of N-boson states which will be discussed in detail in the following section. 

According to Eq.~(\ref{Hamiltonian}), the quantum corrections, with respect to the Hatree-Fock energy, effectively scale as

\begin{equation}
\tilde{g}{(\lambda)}/{({\tilde{g}N})^{\epsilon/2}} \ll 1,
\label{hf-scale}
\end{equation}

\noindent and thus are strongly suppressed in the limit of large $N$~\cite{diagrammatic, Coleman73} or at short distances when $\tilde{g}(\lambda)$ is much less than unity. This is equivalent to the ratio of $\delta \mathcal{H}$ to $\mathcal{H}$.  Using Eq.~(\ref{Hamiltonian}) in the limit of large $N$, or small $\lambda$, we can approximate the Hamiltonian by $V(\lambda) \sim N^2 / (\lambda^d a^{2-d})$. One can then show

\begin{equation}
\frac{\delta \mathcal{H}}{\mathcal{H}} \sim \left( \frac{\lambda^d}{N a^d} \right)^{\epsilon / 2} \ll 1.
\label{lhy-scale}
\end{equation}

\noindent When $d=3$, or $\epsilon = -1$, we recover the Lee-Huang-Yang correction, $\frac{\delta \mathcal{H}}{\mathcal{H}} \sim \sqrt{n a^3}$, where $n$ is the density of the system. This statement also reflects the difficulty with applying our quantum field theory approach to high dimensions; the limit of small $\lambda$ is the strongly interacting regime. The situation is reversed for $d<2$. In this case, the fact that $\epsilon > 0$ implies that the quantum corrections are minimal at short distances or large $N$, as expected.

 $\delta \mathcal{H}$, as mentioned previously, has an imaginary part and the resultant Hamiltonian has a small non-Hermitian component. As a result the dynamics of the scaling parameter $\lambda$, which is further coupled to the microscopic fluctuations of phonons, differs from the usual quantum dynamics of an individual particle. Eq.~(\ref{Hamiltonian}) dictates the energetics of the quantized motion of $\lambda$ which we are exploring. To simplify the discussions, at the moment we neglect the small quantum correction $\delta \mathcal{H}$; we will return to discuss its effect when addressing the issue of lifetimes. The eigenstates of Eq.~(\ref{Hamiltonian}) can be studied by applying the WKB method. The eigenvalues are given by the following equation

\begin{eqnarray}
(n+\frac{1}{2}) \pi&=&\int^{{\lambda}_M}_{\lambda_m} d\lambda \sqrt{2N C_1 (E_n-V(\lambda))}, \nonumber \\
V(\lambda)&=&\frac{C_2N}{2\lambda^2}-C_3\frac{N^2}{\lambda^2}\epsilon (\frac{\lambda}{a})^\epsilon, n=0,1,2...
\label{WKB}
\end{eqnarray}

For a state with energy $E_n$, $\lambda_{M,m}$ represent two classical turning points in the potential $V(\lambda)$. The ground state and low-lying excitations can be easily obtained by further expanding the potential energy in the Hamiltonian near the classical equilibrium position $\lambda_e$,

\begin{equation}
\frac{\lambda_e}{a}=[\frac{C_2}{N C_3}\frac{1}{(2-\epsilon)\epsilon}]^{1/\epsilon}.
\label{ce}
\end{equation}

\noindent Following Appendix C, one can show that 

\begin{eqnarray}
E_n &=&-\frac{C_3^{2/\epsilon}}{2 a^2 C_2^{2/\epsilon-1}}\frac{\epsilon^{1+2/\epsilon}}{(2-\epsilon)^{1-2/\epsilon}}N^{2/\epsilon+1} \nonumber \\
&+& \frac{1}{a^2}\frac{C_3^{2/\epsilon}}{C_1^{1/2} C_2^{2/\epsilon-1/2}}\epsilon^{2/\epsilon+1/2}(2-\epsilon)^{2/\epsilon}N^{2/\epsilon} (n+\frac{1}{2})+...  \nonumber\\
\label{lowE}
\end{eqnarray}

\noindent which is valid when $n \ll N$.

For high energy states with $\lambda_M \gg \lambda_m$ or $n \gg N$, we find that 

\begin{equation}
E_n=\frac{C_3}{a^2} (\frac{2C_1 C_3 f_1^2}{\pi^2})^{(2-\epsilon)/\epsilon}\epsilon^{({2\epsilon-2})/{\epsilon}}N^{{(6-\epsilon)}/{\epsilon}} {(n+\frac{1}{2})^{{2(\epsilon-2)}/{\epsilon}}},
\label{highE}
\end{equation}

\noindent where $f_1(\epsilon)$ is a smooth function of $\epsilon$ near $\epsilon=0$ and is defined in Appendix C. To estimate the number of bound states $n_{max}$, we set $\lambda_M \sim a$.  Since $\lambda_m \ll a$, we find that

\begin{equation}
n_{max}\sim \frac{N^{3/2}}{\epsilon^{1/2}}.
\end{equation}

\subsection{Special cases: $D=1$ and $D=2$}

According to Eq.~(\ref{lowE}), in 1D the ground state energy $E_{n=0}$ has two parts. The leading contribution which is proportional to $N^3$ comes from the potential energy at the equilibrium position $\lambda_e$; the next order contribution, which is proportional to $N^2$, represents the positive shift due to the zero point motion. This scaling behaviour is fully consistent with the McGuire solution of the ground state in 1D~\cite{McGuire64}. In 1D, one can also further map this problem onto the hydrogen atom with high angular momentum as shown in Appendix D. Its spectrum can be expressed as

\begin{equation}
E_n=-\frac{ C_1 C_3^2 N^5}{2 a^2 (n+N\sqrt{C_1 C_2})^2}.
\label{1DE}
\end{equation}

\noindent In the limits of $n\ll N$ and $n\gg N$ Eq.~(\ref{1DE}) is fully consistent with the results of the previous subsection, when $\epsilon$ is taken to be unity.

It is also possible to extend this analysis to d=2 with minor modifications. In 2D, $V(\lambda)$ in  Eq.~(\ref{WKB}) should be

\begin{equation}
V(\lambda) = \frac{C_2 N}{2\lambda^2}-C_3\frac{N^2}{\lambda^2}\frac{1}{\log \frac{a}{\lambda}}.
\label{2dpotential}
\end{equation}

\noindent This result is beyond the usual Gross-Pitaevski equation where the interaction is treated as a constant over all energy scales. In this work the interaction is treated using a running coupling constant which leads to the following quantized solitonic behaviour for N-boson states.

The corresponding ground state and low-lying excitations, following Appendix E, are

\begin{equation}
E_n= -\frac{\exp(2 h_N)}{a^2} (\frac{C_2^2}{4 C_3}-[\frac{C_2^2}{2C_1C_3}]^{1/2}\frac{n+1/2}{N^{1/2}}+...),
\label{lowE2} 
\end{equation}

\noindent where $h_N=\frac{2N C_3}{C_2}$; valid when $n \ll N^{1/2}$.

For the ground state, or low-lying excitations, the first term again is the potential energy at the classical equilibrium point $\lambda_e$. A similar semiclassical result, which was identified as a self-bound droplet state, was suggested previously in Ref.~\cite{Hammer04}. Here we show that the leading quantum correction due to the zero point motion is proportional to $(n+1/2)/\sqrt{N}$. When $\lambda_M \gg \lambda_m$ or $n \gg N^{1/2}$,

\begin{equation}
E_n= - \frac{C_2 N \exp(2 h_N \sin^2\theta_n)}{2 a^2} {\cot^2\theta_n},
\label{theta}
\end{equation}

\noindent where $\theta_n$ is a solution to the following equation,

\begin{eqnarray}
2N^2 \frac{C_1^{1/2} C_3}{C_2^{1/2}}[\frac{\pi}{2}-\theta_n-\frac{\sin2\theta_n}{2}-\frac{C_2}{8NC_3}\cot \theta_n]=(n+\frac{1}{2})\pi.
\nonumber \\
\label{theta1}
\end{eqnarray}

\noindent For $N^2 \gg  n \gg N^{1/2}$, we find the following solution

\begin{eqnarray}
E_n &=&-\frac{(C^2_2 3\pi)^{2/3}}{2^{5/3} a^2C_1^{1/3}C_3^{2/3}}(\frac{n+\frac{1}{2}}{N^{1/2}})^{2/3} \exp(2h_N) \times \nonumber \\
& &\exp[-2^{4/3}\frac{C_3^{1/3} (3\pi)^{2/3}}{C_1^{1/3}C_2^{1/3}}(\frac{n+\frac{1}{2}}{N^{1/2}})^{2/3}].
\end{eqnarray}

\noindent Inspection of Eq.~(\ref{theta1}) shows the number of bound states is on the order of:

\begin{eqnarray}
n_{max}\sim N^2.
\end{eqnarray}

\subsection{Generalized Tjon Relation}

The general correlation between four-body and three-body bound states were
originally noticed and emphasized by Tjon in Ref.~\cite{Tjon75}. Generalization of these correlations to more than four particles in 3D was also examined numerically in Ref.~\cite{Hanna06, Kievsky27}. Tjon relation points out a simple linear correlation between four- and three-body ground state energies, with its linear slope being close to five under certain conditions. Following the results obtained in the previous section, we conclude that in $d=2- \epsilon$, $E_0(N+1)/E_0(N)$ approaches unity in the limit of large N. And in $d=2$, $E_0(N+1)/E_0(N) $ approaches the value of $\exp(4 C_3/C_2)$.

\section{Effect of Quantum Fluctuations}

In the previous section, the semiclassical motion is given by quantizing the Hamiltonian $\mathcal{H}$, Eq.~(\ref{Hamiltonian}) when $\delta \mathcal{H}$ is neglected. In order to quantize this model we impose canonical quantization conditions, Eq.~(\ref{canonical_quantization}). The result of which produces the discrete spectrum evaluated in the previous section. However, this is not the only quantum effects present in the system. The inclusion of $\delta \mathcal{H}$ yields two additional distinct effects. The real part of $\delta \mathcal{H}$ adds a contribution to the canonical quantization, and alters the previously found spectrum. Whereas the imaginary component of $\delta \mathcal{H}$ gives an estimate of the lifetime of these states. 

The evaluation of $Re \delta H$ from Eq.~(\ref{Hamiltonian}) for finite $\epsilon$ yields:

\begin{equation}
Re \delta \mathcal{H} = \frac{\bar{C}(d) N^{1 + d/2}}{a^2} \left( \frac{a}{\lambda} \right)^{d^2/2} + \frac{C}{\lambda^2} - \frac{C' N}{a^{\epsilon} \lambda^d}.
\label{eq:real_delta_H}
\end{equation}

\noindent When evaluating the real part of Eq.~(\ref{Hamiltonian}), it is found that $Re \delta H$ has contributions from fluctuations with wavelengths on the order of $\lambda$. As mentioned previously, the spectrum of the fluctuations at this scale is dependent upon the boundary conditions of the system. Therefore the explicit form of Eq.~(\ref{eq:real_delta_H}) is primarily valid for a finite box with periodic boundary conditions. 

In this approximation, if $\bar{C}(d)$ is equal to zero, the effect of the real part of $\delta \mathcal{H}$ will be to renormalize the original semiclassical Hamiltonian. That is one can replace $C_2$ and $C_3$ with renormalized coefficients, which is the case for $d=1$. For arbitrary $\epsilon$, $\bar{C}(2-\epsilon)$ is non zero, and thus modifies the original Hamiltonian by adding an additional term. This term, as well as the corrections to the renormalized coefficients $C_2$, and $C_3$, are suppressed for large $N$, and can be neglected in our analysis.

\section{Lifetime of N-boson Excited States}

The second affect of $\delta \mathcal{H}$ is to imbue the semiclassical solutions with a finite lifetime. Since these states are collective and coupled with a large numbers of phonons, they will live for a finite lifetime before emitting low energy phonons and undergo a subsequent decay. To estimate the lifetime, we now switch on the quantum corrections and include $\delta \mathcal{H}$ in our analysis. The effective potential in Eq.~(\ref{WKB}) then has a small imaginary term from $\delta \mathcal{H}$. Assuming the eigenvalue now has a small imaginary part $i \delta E_n$ and applying the WKB approach for the eigenvalue problem shown above, we find the lifetime of high-lying excitations (${\tau(E_n)}=1/\delta E_n$) should be given as

\begin{eqnarray}
\frac{1}{\tau(E_n)}&=& \frac{\int d\lambda  {Im} \delta \mathcal{H} [{2(E_n-V(\lambda))}]^{-1/2} }{\int d\lambda [2(E_n-V(\lambda))]^{-1/2}}.
\label{lifetime}
\end{eqnarray}

\noindent The details leading to Eq.~(\ref{lifetime}) are given in Appendix F. Note $d\lambda/\sqrt{2m(E-V(\lambda))}$ measures the time duration within the window $[\lambda, \lambda+d\lambda]$. Thus Eq.~(\ref{lifetime}) is equivalent to a semi-classical average of ${Im} \delta \mathcal{H}$ for a state with eigenvalue $E_n$. Following Eq.~(\ref{lifetime}), we obtain the scaling behaviour of the life time,

\begin{eqnarray}
\frac{1}{\tau(E)} &=& \frac{C_4}{2a^2 C_3^{1-\epsilon/2}} N^{\epsilon/2}(Ea^2)^{1-\epsilon/2} f_2(\epsilon), \nonumber \\
\end{eqnarray} 

\noindent where $f_2(\epsilon)$ is a smooth function defined in Appendix F. One can thus see that the lifetime is also a smooth function near $\epsilon=0$. In particular, upon evaluation one finds that the lifetimes for 1D and 2D are:

\begin{eqnarray}
\tau_{1D} (E_n) &=& \frac{3 \pi^{1/2} C_3^{1/2} a}{C_4 \ 2^{7/2}} \frac{1}{\sqrt{|E_n| N}}, \\
\tau_{2D} (E_n) &=& \frac{6 C_3}{\pi^3 C_4} \frac{1}{|E_n|}.
\label{eq:lifetimes_explicit}
\end{eqnarray}

\noindent If one substitutes the expressions for the spectrum of the high-lying excitations, Eqs.~(\ref{highE}) and (\ref{theta}), the result simplifies to:

\begin{eqnarray}
\tau_{1D} (E_n) &=& \frac{3 \pi^{1/2} a^2}{8 C_4} \frac{1}{\sqrt{C_1 C_3}} \frac{\left( n + \frac{1}{2} \right) }{N^3}, \\
\tau_{2D} (E_n) &=& \frac{12 C_3}{\pi^3 C_2 C_4} \frac{a^2}{N} e^{-2 h_N \sin^2 \theta_n} \tan^2(\theta_n),
\label{eq:lifetimes_high_energies}
\end{eqnarray} 

\noindent Where in defining the 2D lifetime in Eq.~(\ref{eq:lifetimes_high_energies}), we use $h_N = \frac{2 C_3 N}{C_2}$ and Eq.~(\ref{theta1}).

\section{Discussion of Energy Landscape}

\begin{figure}
\begin{center}
\includegraphics[width=0.5\textwidth]{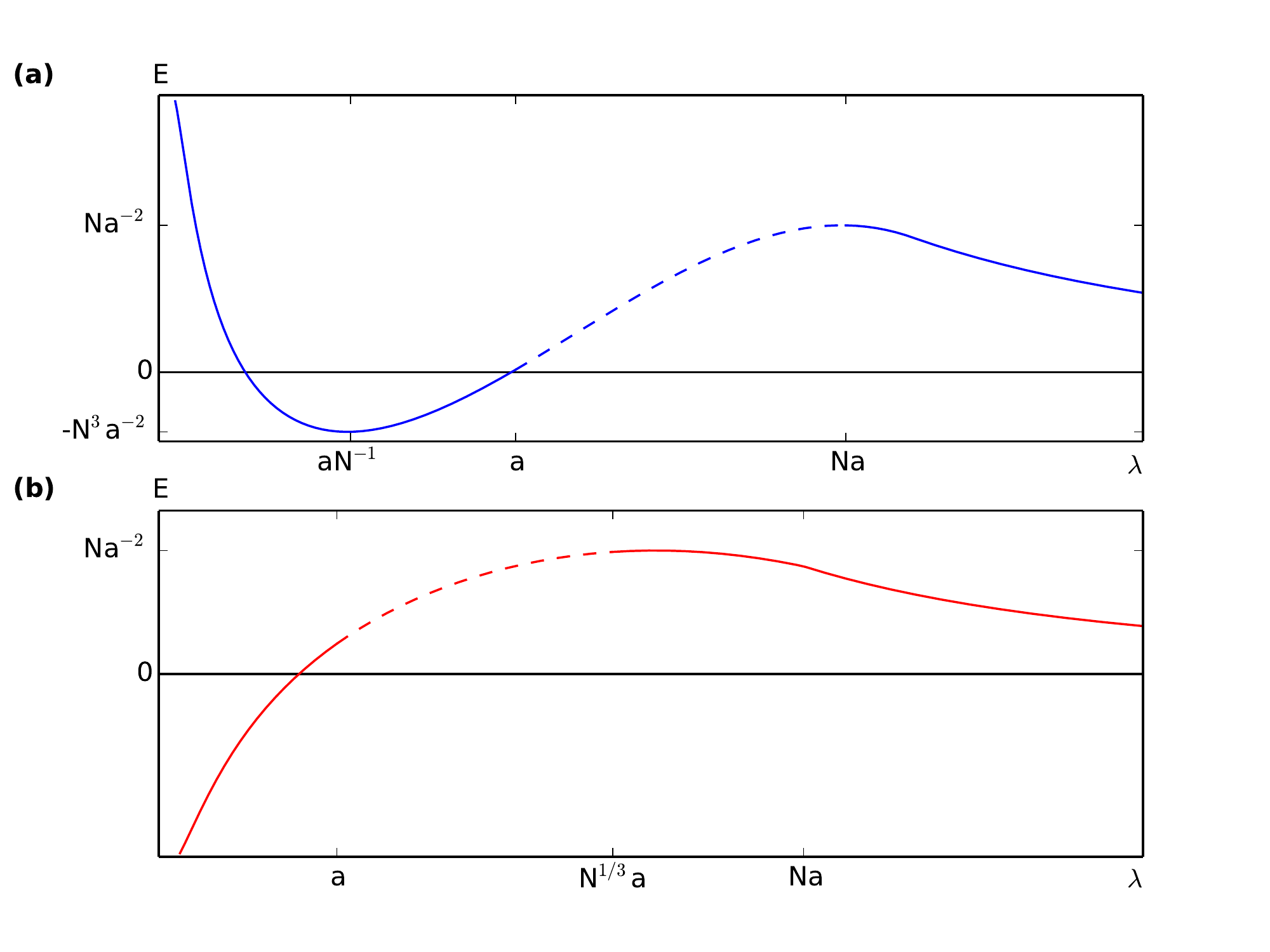}
\caption{The variational energies for 1D and 3D atomic droplets of size $\lambda$ are shown in parts $(a)$ and $(b)$, respectively. In 1D, For $\lambda \ll a$, $E \sim N / \lambda^2$, and for $Na \ll \lambda$, $E \sim N^3 / \lambda^2$. The $N$-boson droplet states found in this work describe the variational energy in the region $a / N < \lambda < a$, where $N$ is the number of particles and $a$ is the scattering length. The energy landscape is unknown in the region $ a < \lambda < Na$, but we propose the form shown in the figure above, given by the dashed line, by analytically matching the results for $0< \lambda <a$ and for $Na< \lambda$. In 3D, the landscape is unknown for $\lambda < N^{1/3}a$. For $N a \ll \lambda$, the gas is weakly interacting and has $E \sim N /\lambda^2$. In the dilute limit, $N^{1/3}a < \lambda < Na$, $E \sim N^2 a / \lambda^3$. Based on the simple scaling analysis, the energy in the semiclassical approximation has to have the form $E \sim -N^2 / \lambda^2$ for $\lambda \ll a$. The form of the variational energy in the non-dilute limit, $a < \lambda < N^{1/3}a$, given by the dashed line, is obtained by extrapolating between the semiclassical energy in the limit $\lambda \ll a$, and the dilute limit.}
\label{fig1}
\end{center}
\end{figure}

This analysis extends the knowledge of the energy landscape for interacting Bose gases with length scales $\lambda < a$.  A plot of the variational energy for 1D and 3D as a function of length, $\lambda$, is shown in parts $(a)$ and $(b)$ of Fig.~(\ref{fig1}), respectively. The variational energy of the system is equivalent to the energy when the system is prepared as a condensate of size $\lambda$. If the gas is not in the condensate phase, the presence of features like deeply bound molecular states will further lower the energy. In this discussion we assume the gas is in the condensate phase for simplicity. In 1D, for asymptotically short length scales, the gas acts as if it were free with energies scaling like $E \sim N / \lambda^2$, since $\tilde{g} \sim \lambda / a$ vanishes asymptotically as $\lambda$ tends to zero. For distances larger than $N a$, a Tonks-Girardeau gas is formed \cite{Tonks60} with energy $E \sim N^3 / \lambda^2$. The droplet states found in our analysis occupy the region where $a / N < \lambda <a$. For $\lambda = a / N$ there exists a minimum which has the semiclassical energy $E = -N^3 / a^2$ (modified by the presence of $\delta \mathcal{H}$). For larger values of $\lambda$, the energy of the droplet states approach zero, with the last state having a length scale $a$, the scattering length. For distances $a < \lambda < N a$, the energy landscape is unknown. We propose that the variational energy must have the form indicated by the dashed line in part $(a)$ of Fig.~(\ref{fig1}). This result is obtained by analytically matching the results known for $\lambda < a$ to those of $Na < \lambda$. This proposed form then has a maximum at $\lambda \approx Na$, with $E \sim N / a^2$.

In 3D, the variational energy is known in three regions: $\lambda \ll a$, $N^{1/3}a < \lambda < Na$, and $Na < \lambda$.  For arbitrarily short distances, $\lambda \ll a$, the attractive interaction dominates the kinetic energy and gives rise to the ultraviolet catastrophe. The exact behaviour at this scale is not known. However, using the simple scaling form of the coupling constant, $\tilde{g}_2$, we posit that the interaction is attractive and scales like the kinetic energy; that is  $E \sim -N^2 / \lambda^2$, since $\tilde{g} = -2 \pi^2$ as $\lambda$ tends to zero. For large distances, $N a \ll \lambda $, the particles are weakly interacting and the total energy still scales like the kinetic energy, $E \sim N / \lambda^2$. In the final region, $N^{1/3}a <  \lambda < Na$, known as the dilute limit, the energy has the form $E \sim N^2 a / \lambda^3$. In the non-dilute limit, where $a < \lambda < N^{1/3}a$, the scaling behaviour of the variational energy is not known. The dashed line shown in part $(b)$ of Fig.~(\ref{fig1}) is obtained by analytically matching the results in the limit $\lambda \ll a$, and the dilute limit. This speculated form of the effective potential has a maximum value of $E \sim N / a^2$ located at $\lambda = N^{1/3} a$.

As seen in part$(b)$ of Fig.~(\ref{fig1}), a speculated instability occurs at the transition from the dilute to non-dilute limit. This instability is consistent with the previous work done in Ref.~\cite{Mashayekhi13}. At this point, the pressure, defined as $P = - \frac{\partial E}{\partial \lambda}$ is zero. When the system passes into the non-dilute limit, the pressure becomes negative; a signature of collapse. Since the pressure is always negative for $\lambda < N^{1/3} a$, the attractive interactions enslave the collapsing dynamics and the resulting variational energy is not bounded from below. Such behaviour is absent in the 1D case, where the variational energy is bounded from below at the semiclassical equilibrium length $\lambda = a /N$.

\section{Conclusions}

To summarize, we have obtained the universal scaling properties of low dimension N-boson solitonic droplet states for length scales $\lambda < a$, where $a$ is the scattering length. The main results are in Eqs.~(\ref{lowE}), (\ref{highE}), (\ref{1DE}), (\ref{lowE2}), (\ref{theta}), (\ref{lifetime}). These results describe the three types of quantum effects acting on the semiclassical solution. The first, the inertial quantization, quantized the overall motion of the system and discretized the energy levels. Quantum corrections to the partition functional yield two additional distinct effects. Firstly, the real part of this correction was shown to alter the semiclassical Hamiltonian by renormalizing the coefficients $C_{1,2,3,4}$ and by adding an additional term. However, these changes are suppressed in the limit of large $N$, which is the focus of our analysis. Secondly, the imaginary part of the quantum correction added an imaginary piece to the energy levels, giving them a finite lifetime.

At last, let us comment on the implications for cold gases currently under study. For a wide class of broad Feshbach resonances studied in many labs where the hyperfine coupling is strong, the effective range of the resonant potentials is much shorter than either the scattering lengths or inter-particle distance of quantum gases. In our work, this criterion is met by stating $\lambda_e > r^*$, or $N \leq \left(\frac{a}{r^*} \right)^{\epsilon}$, where $\lambda_e$ is the minimum of the potential, and $r^*$ is the effective range. For this reason, we can characterize near resonance physics using an attractive contact interaction and our result here can be applied to a wide range of isotopes.

Our analysis on scaling properties of quantum states also has many consequences on the dynamics of quantum gases. For a Bose gas initially prepared in a condensate, but further subject to a zero-range attractive inter-atomic interaction at a later time $t$, the ultimate fate of such a quantum gas is dictated by short distance asymptotic behaviours of scale dependent coupling constants. The dynamics of the gas will fall into two classes: 1) {\it asymptotically free collapsing} where all N-body interactions become irrelevant at short distances; 2) {\it asymptotically subjugated collapsing} where at short distances all interactions become dominating enslaving the collapsing dynamics.  

The investigation of collapsing dynamics is becoming accessible to experiment; one of the most prominent experiment being performed by Carl Wieman's group \cite{Wieman01}. The result of these collapses lead to several interesting phenomena, including the appearance of Bose-novas and jets, and the formation of bright matter-wave solitons \cite{Carr00, Garcia98, Wieman01, Strecker02, Khaykovich02} and dark solitons \cite{Burger99, Denschlag00}. In particular, the formation of solitons is preceded by a collapse during which many of the atoms initially in the condensate are lost. The dynamics describing the initial condensate to the formation of the observed bright matter wave soliton trains will be examined in the future.

We want to thank NSERC(Canada) and Canadian Institute for Advanced Research for their lasting support. One of the authors (M.M.) was also in part supported by an USRA grant from NSERC. We thank Ian Affleck, and Shina Tan for useful discussions. Fei Zhou also wants to thank INT, University of Washington for its hospitality during the workshop on "Universality in few-body systems" in May, 2014.

\appendix

\numberwithin{equation}{section}
\renewcommand\theequation{\Alph{section}.\arabic{equation}}

\section{Lagrangian Formalism}

The Lagrangian density for a bosonic field, $\psi$, with mass $m$, and contact interactions, $g_2$, is given by:

\begin{equation}
\mathcal{L} = i \hbar \psi^* \partial_t \psi - \psi^* \left( -\frac{\hbar^2 \nabla^2}{2 m } \right) \psi -\frac{g_2}{2} \psi^* \ \psi^* \ \psi \ \psi,
\label{eq:lagrangian_density}
\end{equation}

\noindent where $g_2$, is defined as:

\begin{equation}
g_2 =- \frac{1}{a^\epsilon}\frac{(4\pi)^{1-\epsilon/2}}{\Gamma(\epsilon/2)}.
\label{eq:g2}
\end{equation}

\noindent From this point forward, we work in units where $\hbar$ and $m$ are set to unity. It is then advantageous to work with two new fields: the density field $\rho$, and the phase field $\phi$, given by $\psi = \sqrt{\rho} e^{i \phi}$. The transformed Lagrangian density is:

\begin{eqnarray}
\mathcal{L} = \left( - \rho \dot{\phi} - \frac{1}{2} \nabla \sqrt{\rho} \cdot \nabla \sqrt{\rho} - \rho \left( \nabla \phi \right)^2 - \frac{g_2}{2} \rho^2 \right) \nonumber \\
+ \frac{i}{2} \left(\partial_t \rho + \nabla \cdot \left( \rho \nabla \phi \right) \right),
\nonumber
\label{eq:lagrangian_density_transformed}
\end{eqnarray} 

\noindent which has classical equations of motion:

\begin{eqnarray}
0 &=& \partial_t \phi - \frac{1}{2 \sqrt{\rho}} \nabla^2 \sqrt{\rho} + \frac{1}{2} \nabla \phi \cdot \nabla \phi + g_2 \rho ,  \nonumber \\
0 &=& \partial_t \rho + \nabla \cdot \left(\rho \nabla \phi \right) .
\label{eq:eom}
\end{eqnarray}

\noindent From the transformed Lagrangian and the resulting equations of motion, it is immediately obvious that the imaginary term in the Lagrangian density is just the continuity equation, and is constrained to be zero. 

The partition functional is solved in the semiclassical approximation; $\rho$ and $\phi$ are split into semiclassical and fluctuation components: $\rho = \rho_{sc} + \rho_{qc}$, and $\phi = \phi_{sc} + \phi_{qc}$, and the partition function is expanded around $\rho_{sc}$ and $\phi_{sc}$ up to quadratic order in the fluctuation fields. The result is: 

\begin{eqnarray}
Z &=&\int \int \int \left[ D\rho_{sc} \right] \left[ D\rho_{qc} \right] \left[ D\phi_{qc} \right] \nonumber \\
& & \exp \left( i \int dt \mathcal{L}_{sc}+
\mathscr{L}_{qc}(\{\rho_{qc}\},\{\phi_{qc}\}) \right), \nonumber \\
\label{eq:A:partition}
\end{eqnarray}

\noindent and the Lagrangians  $\mathcal{L}_{sc,qc}$ are given as

\begin{eqnarray}
\mathcal{L}_{sc}&=&-\int d{\bf r} \big(\rho_{sc} \partial_t \phi_{sc} +\frac{1}{2}\nabla \sqrt{\rho_{sc}} \cdot \nabla \sqrt{\rho_{sc}}  \nonumber \\
&+&\frac{\rho_{sc}}{2}\nabla \phi_{sc} \cdot \nabla \phi_{sc}  +\frac{g_2}{2} \rho_{sc}\rho_{sc} \big), \nonumber \\
\mathcal{L}_{qc} &=&-\int d{\bf r} \big(\rho_{qc} \partial_t \phi_{qc} +\frac{1}{8}\nabla \frac{\rho_{qc}}{\sqrt{\rho_{sc}}} \cdot \nabla \frac{\rho_{qc}}{\sqrt{\rho_{sc}}}  \nonumber \\
&+&\frac{\rho_{sc}}{2}\nabla \phi_{qc} \cdot \nabla \phi_{qc}  +\frac{g_2}{2} \rho_{qc}\rho_{qc} + {\rho_{qc}}\nabla \phi_{qc} \cdot \nabla \phi_{sc}  \big)+... \nonumber \\
\label{Lqc}
\end{eqnarray}

\noindent Eq.~(\ref{Lqc}), only terms up to quadratic order in the fluctuations are kept, due to the asymptotic freedom present in low dimensions. These Lagrangian densities hold for both a uniform and non-uniform systems.

We now propose the following isotropic scaling ansatz for $\rho_{sc}$:

\begin{equation}
\rho_{sc}(r,t) = \frac{N}{\lambda^d} f \left( \frac{r}{\lambda(t)} \right).
\label{eq:semiclassical_rho}
\end{equation}

\noindent To obtain $\phi_{sc}$, one simply places Eq.~(\ref{eq:semiclassical_rho}) into the second line of Eq.~(\ref{eq:eom}) to obtain:

\begin{equation}
\phi_{sc} = \frac{r^2}{2} \frac{\dot{\lambda}}{\lambda} + \phi_0(t). \nonumber
\end{equation}

\noindent $\phi_0(t)$ is a constant of integration that can be fixed using the other equation of motion. The total result is:

\begin{equation}
\phi_{sc}(r,t) = \frac{r^2}{2} \frac{\dot{\lambda}}{\lambda} + \int dt \frac{B_1}{2 \lambda^2} -\frac{N B_2}{\lambda^d},
\label{eq:semiclassical_phi}
\end{equation}

\noindent where $B_1$ and $B_2$ depend on the specific form of the scaling function $f$, namely $B_1=\frac{1}{4f(0)} \partial^2_x f({x})|_{x=0}$ and $ B_2=g_2 f(0)$.

Before determining the transformed partition function, one must first take care to ensure that the original periodic boundary conditions imposed on the bosonic fields, $\psi$ and $\psi^*$, are encoded in the density and phase fields. The original boundary conditions are:

\begin{equation}
\psi(\vec{r},0) = \psi(\vec{r},\infty ). \nonumber
\end{equation}

\noindent When the semiclassical solutions are placed back into Eq.~(\ref{eq:lagrangian_density_transformed}), there is a term in the action of the from:

\begin{equation}
\int d {\bf r} dt \rho \partial_t \phi. \nonumber
\end{equation}

\noindent Using the fact that the density is normalized at each moment in time, one can easily show:

\begin{equation}
\int d {\bf r} dt \rho \partial_t \phi = \int d {\bf r} dt \rho \partial_t \tilde{\phi} + N \left( \phi_0( \infty ) - \phi_0(0) \right), \nonumber
\end{equation}

\noindent where $\tilde{\phi} = \frac{r^2}{2} \frac{\dot{\lambda}}{\lambda}$. The initial conditions on the phase field acts as a Lagrangian multiplier fixing the number of particles. The initial conditions then do not enter into the dynamics of the system, but merely add an overall phase factor. 

With this in mind, the transformed partition function and Lagrangians are found to be:

\begin{eqnarray}
& & Z \approx \int D\lambda \exp(i\int dt \mathcal{L}_{eff}), \nonumber \\
& &\mathcal{L}_{eff}(\lambda(t)) = \mathcal{L}_{sc}+\mathcal{L}_{qc}\nonumber \\
& &\mathcal{L}_{sc} = C_1 \frac{N}{2} (\partial_t {\lambda})^2 -C_2 \frac{N}{2\lambda^2}+C_3 \epsilon \frac{N^2}{\lambda^{2-\epsilon}a^\epsilon} \nonumber\\
& & \mathcal{L}_{qc} = \frac{1}{2}\sum_{k > 2\pi/\lambda} (\epsilon^2_k + 2 \epsilon_k \mu)^{1/2}-({\epsilon_k+\mu}) \ \ \ \ \ \ \
\label{appendix:qc}
\end{eqnarray}

\noindent where  $\mu = - N/\lambda^{2-\epsilon} {g}_2$. The discussions of $\mathcal{L}_{qc}$ and the evaluation of the functional integrals over the fluctuations are found in Appendix B.

The coefficients appearing in Eq.~(\ref{qc}), are determined by the explicit scaling ansatz used: 

\begin{eqnarray}
C_1 &=& \int d^d x f(x) x^2, \nonumber \\
C_2 &=&  \frac{1}{4} \int d^dx \frac{\nabla f(x) \cdot \nabla f(x)}{f(x)}, \nonumber \\
C_3 &=& \frac{(4 \pi)^{d/2}}{2 \epsilon \Gamma \left( \frac{\epsilon}{2} \right)}   \int d^dx f(x)^2.
\label{eq:coefficients}
\end{eqnarray}

\noindent These coefficients have been calculated for several different scaling ansatz; the results of which are found in Table~(\ref{tab:1d}) and Table~(\ref{tab:2d}).

\begin{table}
\begin{center}
\begin{tabular}{|c|c|c|c|}
\hline
Function & $C_1$ & $C_2$ & $C_3$ \\
\hline
$e^{-x^2}$ & $\frac{1}{2}$ & $\frac{1}{2}$ & $\frac{1}{\sqrt{2 \pi}}$ \\
\hline
$e^{-|x|}$ & $2$ & $\frac{1}{4}$ & $\frac{1}{4}$ \\
\hline
$\cosh^{-1}(x)$ & $\frac{\pi^2}{4}$ & $\frac{1}{8}$ & $\frac{2}{\pi^2}$ \\
\hline
\end{tabular}
\end{center}
\caption{Coefficients for various scaling ansatz $f(x)$, in 1D. These functions are not normalized. The normalization condition is given by $\int d^d x f(x) = 1$.}
\label{tab:1d}
\end{table}

\begin{table}
\begin{center}
\begin{tabular}{|c|c|c|c|}
\hline
Function & $C_1$ & $C_2$ & $C_3$ \\
\hline
$e^{-x^2}$ & $\frac{\sqrt{\pi}}{2}$ & $\sqrt{\pi}$ & $\sqrt{2}$ \\
\hline
$e^{-x}$ & $\frac{\sqrt{\pi}}{4}$ & $\frac{1}{4}$ & $\frac{1}{2 \sqrt{\pi}}$ \\
\hline
$\cosh^{-1}(x)$ & $\frac{\pi^3}{16 G}$ & $\frac{\pi}{32 G}$ & $\frac{1}{4 G^2 \sqrt{\pi}}$ \\
\hline
\end{tabular}
\end{center}
\caption{Coefficients for various scaling ansatz $f(x)$, in 2d. These functions are not normalized. The normalization condition is given by $\int d^d x f(x) = 1$. In this table, $G$ is Catalan's constant $G \approx 0.9160$}
\label{tab:2d}
\end{table}

\section{Evaluation of the quantum correction}
In this appendix, the quantum correction to the Hamiltonian is evaluated. This can be done analytically as $\mathcal{L}_{qc}$ is quadratic in the fluctuations. This procedure is equivalent to the one loop diagrammatic expansion. 

As a first example, consider a uniform translationally invariant system.  In this case, it is most convenient to expand the fluctuations in terms of Fourier modes with wave vectors $\vec{k}$ and frequency $\omega$. The functional integral over the fluctuations, $Z_{qc}$, becomes:

\begin{eqnarray}
Z_{qc} &=& \int \left[D \rho_{qc} \right] \left[ D \phi_{qc} \right] \prod
_{k, \omega}  \exp~\left[ - \omega \rho_{qc,+} \phi_{qc,-}  \right. \nonumber \\
&-& \left. i \frac{k^2}{8 \rho_0} \rho_{qc,+} \rho_{qc,-} - i \frac{1}{2} \rho_0 k^2 \phi_{qc,+} \phi_{qc,-} \right. \nonumber \\
&-& \left. i\frac{g_2}{2}  \rho_{qc,+} \rho_{qc,-} \right]. \ 
\label{eq:quantum_partition_function}
\end{eqnarray}

\noindent $\rho_{qc,\pm}$ and $\phi_{qc,\pm}$ in the above equation represent $\rho_{qc}(\pm \vec{k}, \pm \omega)$, and $\phi_{qc}(\pm \vec{k}, \pm \omega)$, respectively. Eq.~(\ref{eq:quantum_partition_function}) is a set of two functional integrals involving an exponential function which is quadratic in $\rho_{qc}$, and $\phi_{qc}$. A functional integral over an exponential which is quadratic in these fields can be viewed as an infinite number of regular Gaussian integrals, one for each value of $\vec{k}$ and $\omega$. In terms of a functional integral, the measure $\left[ D \rho_{qc} \right]$ and $\left[ D \phi_{qc} \right]$ are defined in such a way as to absorb the constants that emerge in integration. To be specific, consider performing the functional integral over $\phi_{qc}$. The measure $\left[ D \phi_{qc} \right]$, is just $\prod_{\vec{k},\omega} d \phi_{qc}(\vec{k}, \omega)$ up to a multiplicative constant which is neglected. The integral to be considered then is:

\begin{eqnarray}
I = \prod_{\vec{k}, \omega} \int_{-\infty}^{\infty} d \phi_{qc}(\vec{k}, \omega) \exp \left[ -i \frac{\rho_0 k^2}{2} \right. \nonumber \\
\left.  \times \left( \phi_{qc,+} \phi_{qc,-} - \frac{2 i \omega}{\rho_0 k^2} \rho_{qc,+} \phi_{qc,-} \right) \right],
\label{eq:Int_1}
\end{eqnarray}

\noindent which is just a product of one dimensional integrals. It is possible to complete the square in the exponential, and use the standard result for a one dimensional Gaussian integral to evaluate Eq.~(\ref{eq:Int_1}).  Upon doing said integration, $Z_{qc}$ can be shown to have the form:

\begin{eqnarray}
Z_{qc} = \prod_{\vec{k}, \omega} \sqrt{\frac{1}{i \rho_0 k^2}} \int_{-\infty}^{\infty} d\rho_{qc}(\vec{k},\omega) exp \left[ -\frac{i}{2 \rho_0 k^2} \right. \nonumber \\
\left. \times \left\lbrace \left(\frac{k^2}{2}\right)^2 + \frac{2g_2 \rho_0 k^2}{2} - \omega^2 \right\rbrace \rho_{qc,+} \rho_{qc,-} \right]. \ \ \ 
\end{eqnarray}

\noindent Performing the same procedure to evaluate the functional integral over $\rho_{qc}$ gives the final result for $Z_{qc}$:

\begin{eqnarray}
Z_{qc} & = & \prod_{\vec{k},\omega} \sqrt{\frac{1}{\omega^2 - \omega_k^2 + i \delta}}. \nonumber \\
\omega_k & = & \sqrt{\left(\epsilon_k\right)^2 + 2 \mu \epsilon_k}
\label{eq:final_quantum_z} 
\end{eqnarray}

\noindent In the above equation, $\epsilon_k = \frac{k^2}{2}$, the energy of a free particle, $\omega_k$ is the Bogoliubov dispersion of phonons in a weakly interacting Bose gas, $\mu$ is the chemical potential given by $\frac{g_2 N}{\lambda^d}$, and the factor of $i \delta$ has been added to ensure convergence. In order to reduce the result of Eq.~(\ref{eq:final_quantum_z}) into the form of a Hamiltonian, we note that  $\delta H = -\frac{1}{i T}  ln(Z_{qc})$, where $T$ is some time parameter. One can show by performing the sum over frequency space that:

\begin{equation}
\delta H =  \frac{1}{2} \sum_{\vec{k}} \omega_k.
\label{eq:final_delta_E}
\end{equation}

For a non-uniform system, it is necessary to understand the form of the semiclassical ansatz, as it will dictate the form of the fluctuations. However, the choice of boundary conditions will only change the low-lying modes of the fluctuation spectrum, which are discrete; the fluctuation modes with $k > \sqrt{\mu}$ will be unaltered. In the limit we are interested in, $N \gg 1$, the summation over the fluctuations will be dominated by the modes with $k \sim \sqrt{\mu}$, much larger than the cut off $\frac{2 \pi}{\lambda}$. Therefore, an approximation to the real fluctuations will only affect the modes which are negligible in our analysis. We then choose to evaluate the functional integral over the fluctuations by assuming a uniform density up to the scaling parameter, $\lambda$. This assumption states that the non-uniformity of the system is slowly varying on length scales less than $\lambda$. Since the system is approximated to have uniform density up to a scale $\lambda$, it is possible to perform a finite Fourier transform and express the density and phase fields in terms of Fourier modes that have $|\vec{k}| \geq \frac{2 \pi}{\lambda}$. This is equivalent to separating the fast motions of the fluctuation fields and the slow motion of the semiclassical fields. The subsequent analysis is the exact same as in the uniform case, except for the sum given in Eq.~(\ref{eq:final_delta_E}) is for all $\vec{k}$, with magnitudes $|k| \geq \frac{2 \pi}{\lambda}$.

Moreover in the uniform and non-uniform cases, the result given by Eq.~(\ref{eq:final_delta_E}) is formally divergent but can be regularized to produce a finite result. To do so it is necessary to introduce terms which eliminate the ultraviolet divergences. Such a term comes naturally from the diagrammatic loop expansion of an interacting Bose gas. The finite correction due to the fluctuations is given by:

\begin{equation}
\delta H =  \frac{1}{2} \sum_{|\vec{k}| > \frac{2 \pi}{\lambda}} \omega_k - \epsilon_k - \mu .
\label{eq:final_delta_E_regularized}
\end{equation}

When the interactions are attractive, $g_2$ and the chemical potential are negative. As a result there is a range of values of $\vec{k}$  for which $\omega_k$ in Eq.~(\ref{eq:final_delta_E}) is imaginary, which implies $\delta H$ is a complex number. The imaginary piece of $\delta H$ arises from phonons with small values of $\vec{k}$, and is always finite; whereas the real piece is divergent and requires regularization. 

In its entirety Eq.~(\ref{eq:final_delta_E_regularized}) can be converted into an integral using density of states:

\begin{equation}
\delta H = C_4 \frac{(\epsilon N)^{2-\epsilon / 2} }{2 a^2} \left( \frac{a}{\lambda}\right)^{(2 - \epsilon)^2 / 2} F(\frac{\lambda}{\pi} \sqrt{\frac{\mu}{2}}),
\label{eq:def_dH}
\end{equation}

\noindent where,

\begin{equation}
F(\eta) = \frac{2 \pi^d}{\Gamma(d/2)} \int_{1 / \eta}^{\infty} dx x^{d-1} \left[ \sqrt{x^4 - 2x^2} - x^2 + 1 \right]. 
\label{eq:def_F}
\end{equation}

\noindent $F(\eta)$ is a smooth function of $\epsilon$ for $\epsilon > 0$. This function quickly approaches its asymptotic form of $-\frac{C}{\lambda}$, for some constant $C$. In the following analysis, it is possible to set $\eta$ equal to infinity, as it will not affect the results greatly. 

In obtaining Eqs.~(\ref{eq:def_dH}) and (\ref{eq:def_F}), we assumed that there was a continuum of states starting from the cut off $k = \frac{2 \pi}{\lambda}$. Around this cut off, the phonon modes are actually discrete; not continuous. However, this can be shown to generate a negligible change to the quantum corrections. To quantify this, we focus on the one-dimensional case; the structure for higher dimensions is more complicated, but is assumed to have the same form. In 1D, we use the Euler-Maclaurin formula for the difference between a summation and a integral when $\frac{2\pi}{\lambda} < k < \sqrt{\mu}$. In order to apply this formula, we first assume that in the region of interest, $\omega_k$ can be approximated by the phonon spectrum $\omega_k = \sqrt{|\mu|} k$. Whether these phonons are stable or not is irrelevant to this discussion.

The Euler-Maclaurin formula states:

\begin{eqnarray}
&& \sum_{\frac{2 \pi}{\lambda}}^{k_0} \omega(k) - \int_{\frac{2 \pi}{\lambda}}^{\sqrt{|\mu |}} \omega(k) \sim \frac{\omega(\sqrt{|\mu |}) + \omega(2 \pi / \lambda)}{2} \nonumber \\
&+& \sum_{l=1}^{\infty} \frac{B_{2l}}{(2l)!} \left( \omega^{(2l-1)}(\sqrt{| \mu |}) - \omega^{(2l-1)}(\frac{2 \pi}{\lambda}) \right). \\ \nonumber
\end{eqnarray}

\noindent where $B_{2l}$ are the Bernoulli numbers. For the given form of $\omega(k)$, the sum in the Euler-Maclaurin formula is equal to zero and thus the difference between the sum and the integral is:

\begin{equation}
\Delta = \sum_{\frac{2 \pi}{\lambda}}^{k_0} \omega(k) - \int_{\frac{2 \pi}{\lambda}}^{\sqrt{|\mu |}} dk \omega(k) \sim \frac{\sqrt{|\mu |}}{2} \left( \sqrt{ | \mu | } + \frac{2 \pi}{\lambda} \right).
\end{equation}

\noindent As a result, to leading order in $N$, we introduce a correction of order $| \mu | \approx N / (a \lambda)$.  The second term proportional to $\sqrt{ | \mu |}$ introduces a correction of order $N^{1/2}$ which, in the large $N$ limit, is negligible. Comparing this to the value of the integral, one obtains:

\begin{eqnarray}
\Delta \left(\int_{\frac{2 \pi}{\lambda}}^{\sqrt{| \mu |}} dk \omega k \right)^{-1}  &\approx & \frac{1}{\sqrt{| \mu |}} \nonumber \\
&=& \frac{a \lambda}{N} \ll 1.
\end{eqnarray}

\noindent where we note that in the area of interest, $\lambda \geq \frac{a}{N}$, and $N \gg 1$, as will be seen in subsequent discussions. This implies that we are free to assume that the fluctuation spectrum is continuous for all wave vectors $\frac{2 \pi}{\lambda} \leq | \vec{k} |$.

It is now possible to determine when the imaginary part of Eq.~(\ref{eq:def_F}) becomes non-zero. Physically this is when the phonon modes become unstable and acquire a finite lifetime. When one examines the structure of $\omega_k$ in Eq.~(\ref{eq:final_quantum_z}), it is found that for $\lambda < \lambda_c$, $\omega_k$ becomes purely real, where:

\begin{equation}
\lambda_c = \left( \frac{\pi}{4} \frac{\left( 4 \pi \right)^{\frac{\epsilon}{2}} \Gamma \left( \frac{\epsilon}{2} \right)}{N} \right)^{\frac{1}{\epsilon}} a.
\label{eq:critical_size}
\end{equation}

\noindent Thus, an instability exists in the phonon modes for large scaling parameters $\lambda$. 

Assuming $\lambda > \lambda_c$, the corresponding imaginary piece of Eq.~(\ref{eq:final_delta_E_regularized}), to leading order, can be written as:

\begin{eqnarray}
Im ( \delta H ) &=& C_4 \frac{(\epsilon N)^{2-\epsilon / 2}}{2 a^2} \left( \frac{a}{\lambda} \right)^{d^2 / 2} \frac{2 \pi^d}{\Gamma \left( 1 - \frac{\epsilon}{2} \right)} \times \nonumber \\
& & \int_{\frac{1}{\eta}}^{\sqrt{2}} dx x^{d-1}\sqrt{2x^2 - x^4},
\label{eq:imaginary_component}
\end{eqnarray}

\noindent where $\eta$ was previously defined as $\eta = \frac{\lambda}{\pi} \sqrt{\frac{\mu}{2}}$.

The real part of Eq.~(\ref{eq:def_F}) is the remaining portion of the integrand and is  finite for $d<2$ dimensions. The discussion of its effects is postponed to Appendix C.

\section{Evaluation of the Energy Spectra when $D=2 -\epsilon$}

In order to determine the energy spectrum for this system, it is necessary to quantize the motion of $\lambda$. First we consider the case when $\delta H$ is set equal to zero, and quantize the motion of the semiclassical Lagrangian in Eq.~(\ref{Lqc}). This is done using the standard canonical quantization procedure and yields:

\begin{equation}
H = \frac{1}{2 C_1 N} P_{\lambda}^2 + \frac{C_2 N}{2 \lambda^2} - \frac{C_3 \epsilon N^2}{\lambda^2} \left( \frac{\lambda}{a} \right)^{\epsilon}.
\label{eq:hamiltonian}
\end{equation}

\noindent where:

\begin{equation}
\left[ \lambda, P_{\lambda} \right] = i \hbar
\end{equation}

\noindent To progress further, it is advantageous to use the WKB formalism which states:

\begin{eqnarray}
\left(n + \frac{1}{2} \right) \pi & = &  \int_{\lambda_m}^{\lambda_M} d \lambda \ \sqrt{2 N C_1 \left( E_n - V(\lambda) \right)}, \nonumber \\
 V(\lambda)& = & \frac{C_2 N}{2 \lambda^2} - \frac{C_3 \epsilon N^2}{\lambda^2} \left( \frac{\lambda}{a} \right)^{\epsilon},
\label{eq:WKB}
\end{eqnarray}

\noindent where $\lambda_m$ and $\lambda_M$ are the classical turning points of the potential given above. Eq.~(\ref{eq:WKB}) can be solved for both the low-lying and high-lying excitations. In the case of low-lying excitations, it is possible to expand the potential to quadratic order around its minimum, located at:

\begin{equation}
\lambda_e = \left( \frac{C_2}{N C_3} \frac{1}{\epsilon (2 - \epsilon)} \right)^{\frac{1}{\epsilon}} a.
\label{eq:equilibrium_lambda}
\end{equation}

\noindent The resulting integral is elementary and yields:

\begin{eqnarray}
E_n &=&-\frac{C_3^{2/\epsilon}}{2 a^2 C_2^{2/\epsilon-1}}\frac{\epsilon^{1+2/\epsilon}}{(2-\epsilon)^{1-2/\epsilon}}N^{2/\epsilon+1} \nonumber \\
&+& \frac{1}{a^2}\frac{C_3^{2/\epsilon}}{C_1^{1/2} C_2^{2/\epsilon-1/2}}\epsilon^{2/\epsilon+1/2}(2-\epsilon)^{2/\epsilon}N^{2/\epsilon} (n+\frac{1}{2})+...  \nonumber\\
\label{eq:lowlying_en}
\end{eqnarray}

\noindent which is valid for $\epsilon \neq 0$ and $n \ll N$. 

For the high-lying states when $\epsilon \neq 0$, it is necessary to bring Eq.~(\ref{eq:WKB}) into a form that can be expanded in powers of some small quantity. To determine this quantity, it can be shown that $\lambda_m$ and $\lambda_e$ are of the same order of magnitude and scale linearly with $N$. This is a consequence of the shape of the potential; the repulsive part of the potential is very steep, $\lambda^{-2}$, whereas the tail of the potential scales like $\lambda^{2-\epsilon}$ for finite $\epsilon$. In the case of high-lying excitations one can expect $\lambda_M \gg \lambda_e \approx \lambda_m$. 

In order to reduce Eq.~(\ref{eq:WKB}) into powers of $\frac{\lambda_e}{\lambda_M}$, one first expresses the total energy, $E$, in terms of the classical turning point $\lambda_M$. This will reduce the integral into the following form:

\begin{eqnarray}
& &(n + \frac{1}{2}) \pi = \sqrt{\frac{2 \epsilon C_1 C_3 N^3 \lambda_M^{\epsilon}}{a^{\epsilon}}} \times \nonumber \\ 
& & \int_{\frac{\lambda_e}{\lambda_M}}^{1} \frac{dx}{x^{1- \epsilon / 2}} \sqrt{1-x^{2 - \epsilon} - \frac{(2- \epsilon)}{2}\frac{\lambda_e}{\lambda_M} \left( 1 - x^2 \right)}. \nonumber \\ 
\label{eq:highlying_en_1}
\end{eqnarray}

\noindent When expanding Eq.~(\ref{eq:highlying_en_1}) in terms of $\frac{\lambda_e}{\lambda_M}$,  the resulting first non-trivial order of $\frac{\lambda_e}{\lambda_M}$ is $ O \left( \frac{\lambda_e}{\lambda_M} \right)^{\epsilon / 2}$, which vanishes for large system sizes. As a result, it is possible to neglect the terms proportional to $\frac{\lambda_e}{\lambda_M}$ in the square root of Eq.~(\ref{eq:highlying_en_1}). The integral then is analytical with the result:

\begin{equation}
E = -\frac{C_3}{a^2} \left( \frac{2 C_1 C_3 f_1^2}{\pi^2} \right)^{\frac{2 - \epsilon}{\epsilon}} \epsilon^{\frac{2 - 2 \epsilon}{\epsilon}} N^{\frac{6 - \epsilon}{\epsilon}} \left( n + \frac{1}{2} \right)^{\frac{2(\epsilon-2)}{\epsilon}},
\label{eq:high_lying_spectrum}
\end{equation}

\noindent where we note that $E \approx -\frac{N^2 \epsilon C_3}{ \lambda_M^{2-\epsilon} a^{\epsilon}}$. The function $f_1$ is defined as:

\begin{equation}
f_1 = (2-\epsilon) \int_0^1 dx x^{1-\epsilon /2} (1-x^{2-\epsilon})^{-1/2}, \nonumber
\end{equation} 

\noindent which is a smooth function of $\epsilon$, near $\epsilon = 0$.

An estimate of the maximum number of bound states, $n_{max}$, can be determined by setting $\lambda_M \approx a$ in Eq.~(\ref{eq:highlying_en_1}). The result is: 

\begin{equation}
n_{max} \sim \frac{N^{3/2}}{\epsilon^{1/2}} 
\end{equation}

Finally, we are now in a position to consider the case when $Re(\delta H)$ is non-zero. Since $Re(\delta H)$ is regularized, the result is finite and given by:

\begin{equation}
Re \delta \mathcal{H} = \frac{\bar{C}(d) N^{1 + d/2}}{a^2} \left( \frac{a}{\lambda} \right)^{d^2/2} + \frac{C}{\lambda^2} - \frac{C' N}{a^{\epsilon} \lambda^d},
\label{eq:realdeltaH}
\end{equation}

\noindent for finite $\epsilon$. This result is dependent on the choice of boundary conditions for the system, as $Re \delta \mathcal{H}$ is dominated by fluctuations with wavelengths on the order of $\lambda$. As mentioned in Appendix B, the form of fluctuations on these length scales depends on the boundary conditions of the system. Therefore the above result holds specifically for the finite box $f$-function.

The last two terms of Eq.~(\ref{eq:realdeltaH}) have the same structure as $V(\lambda)$ in the semiclassical Hamiltonian and merely renormalize the coefficients $C_2$ and $C_3$. It is then possible to rewrite the total Hamiltonian as:

\begin{equation}
H = \frac{1}{2 C_1 N} P_{\lambda}^2 + \frac{\bar{C}_2 N}{2 \lambda^2} - \frac{\bar{C}_3 N^2}{\lambda a} + \frac{\bar{C}(d) N^{1 + d/2}}{a^2} \left( \frac{a}{\lambda} \right)^{d^2/2}.
\end{equation}

\noindent The presence of the additional term states that the effect of $Re \delta \mathcal{H}$, is not to simply renormalize the Hamiltonian. However, the corrections to the semiclassical Hamiltonian are suppressed in the limit of large $N$, which is the limit of interest in our analysis. Thus these contributions can be neglected.

\section{Alternative Solution for 1D}

In one dimension, the Lagrangian has a form similar to a hydrogen atom and the resulting quantum mechanical system can be solved in a manner similar. We start with the 1D Lagrangian given by:

\begin{equation}
L = \frac{1}{2} N C_1 \left( \dot{\lambda} \right)^2 - \frac{N}{2 \lambda^2} + \frac{ N^2}{\lambda a} C_3, \nonumber
\end{equation}

\noindent which leads to the following quantized Hamiltonian;

\begin{equation}
H = -\frac{1}{2 N C_1} \frac{\partial^2 }{\partial \lambda^2} + \frac{N C_2}{2 \lambda^2} - \frac{N^2 C_3}{\lambda a}. 
\label{eq:B:hamiltonian}
\end{equation}

For $\lambda \gg 1$, the dynamics are that of a free particle, and one can write $\psi(\rho) = e^{- \rho}$, where $\rho = \sqrt{-2N C_1 E} \lambda = \kappa \lambda$, and $E$ is the energy of the bound state.  However, for $\lambda \ll 1$, the solution has the form $\psi(\rho) = \rho^k$, where $k$ satisfies $k(k-1) = N^2 C_1 C_2$, that is:

\begin{equation}
k = \frac{1}{2} + \sqrt{\frac{1}{4}+ N^2 C_1 C_2}.
\label{eq:B:k_def}
\end{equation}

\noindent Combining these forms, the general solution to the Hamiltonian is:

\begin{equation}
\psi(\rho) = e^{-\rho} \rho^k \sum_{n=0}C_n \rho^n.
\end{equation}

When one substitutes this into the time independent Schroedinger equation, governed by the Hamiltonian in Eq.~(\ref{eq:B:hamiltonian}), the following recursion relation is obtained:

\begin{equation}
\frac{C_{n+1}}{C_n} = 2 \frac{(k+u) - \frac{N^3 C_1 C_3 }{\kappa a }}{(k+n+1)(k+n) - N^2 C_1 C_2}.
\label{eq:B:recursion}
\end{equation} 

\noindent The wave function will converge if, for some $n$, Eq.~(\ref{eq:B:recursion}) is zero, that is:

\begin{equation}
n+k = \frac{N^3 C_1 C_3}{\kappa a}.
\end{equation}

\noindent This equation can be used to determine the energy of the bound state, yielding:

\begin{equation}
E = - \frac{1}{2} \frac{C_1 C_3^2 N^5}{a^2 (n + \frac{1}{2} +  \sqrt{\frac{1}{4}+C_1 C_2 N})^2}.
\end{equation}

\noindent The above spectrum simplifies for $N \gg 1$:

\begin{equation}
E = - \frac{1}{2} \frac{C_1 C_3^2 N^5}{a^2 (n + \sqrt{C_1 C_2} N)^2},
\end{equation}

\noindent which looks like the energy spectrum of a Hydrogen atom.

\section{2D Spectrum} 

The same analysis used in Appendix C can be extended to the case $d=2$. The only adjustment is that the potential to be used is:

\begin{equation}
V(\lambda) = \frac{C_2 N}{2 \lambda^2} - C_3 \frac{N^2}{\lambda^2} \frac{1}{log\left(\frac{a}{\lambda} \right)},
\label{eq:2d_potential}
\end{equation}

\noindent which is the $\epsilon \rightarrow 0$ limit of the potential in Eq.~(\ref{eq:WKB}). The calculation of the droplet states around the semiclassical minimum proceeds just as in the 1D case, except for the use of Eq.~(\ref{eq:2d_potential}) for $V(\lambda)$. To begin this analysis, consider evaluating the minimum of Eq.~(\ref{eq:2d_potential}). Setting the first derivative equal to zero yields:

\begin{equation}
0 = 1 - \frac{h_N}{log \left( \frac{a}{\lambda_e}\right)} + \frac{2 h_N}{log^2 \left( \frac{a}{\lambda} \right)},
\label{eq:cancellation}
\end{equation}

\noindent where $h_N = \frac{2C_3 N}{C_2}$. Assuming that $a \gg \lambda_e$, the following result is obtained:

\begin{equation}
\lambda_e = a e^{-h_N}. \nonumber
\end{equation}

\noindent The importance of Eq.~(\ref{eq:cancellation}) will be noted when calculating the high-lying spectrum. At this point, using the above expression for $\lambda_e$, the calculation of the spectrum for the low-lying states is exactly equivalent to the finite $\epsilon$ case. The result is given by:

\begin{equation}
E_n = \frac{e^{2 h_N}}{a^2} \left( -\frac{C_2^2}{2 C_3} + \sqrt{\frac{C_2^2}{2 C_1 C_3}} \frac{\left(n + \frac{1}{2} \right)}{\sqrt{N}} \right),
\label{eq:low_loying_spectrum_2d}
\end{equation}

\noindent  which is valid for $n \ll N^{1/2}$.

For high lying states, one assumes $n \gg N^{1/2}$. In this regime, $a \gg \lambda_M \gg \lambda_e$, and thus the small parameter to be considered is still $\frac{\lambda_e}{\lambda_M}$. To leading order, the WKB formalism states:

\begin{equation}
\left( n + \frac{1}{2} \right) \pi = \sqrt{C_1 C_2 N^3} \int_{\lambda_m}^{\lambda_M} \frac{d\lambda}{\lambda} \sqrt{\frac{2 C_3}{C_2 log(a / \lambda)} - \frac{1}{N}}.
\label{eq:2D_leading_order}
\end{equation}

\noindent However, unlike the finite $\epsilon$ case, the contribution from $\lambda_m$ to the above integral is non-trivial. When $\lambda = \lambda_m$, Eq.~(\ref{eq:cancellation}) can be used to show that the integrand is of order $N^{1/2}$, which is finite in comparison to the contribution to $\lambda_M$. With this in mind, the integral can be performed analytically and the high lying spectrum has the following form:

\begin{equation}
E_n = -\frac{C_2 N}{2 a^2} \cot^2 (\theta_n) e^{2 h_N \sin^2(\theta_n)},
\label{eq:high_lying_spectrum_2d}
\end{equation}

\noindent where $\theta_N$ is a solution to:

\begin{eqnarray}
\left( n + \frac{1}{2} \right) \pi &=& \frac{2 N^2 C_1^{1/2}}{C_2^{1/2}} \left[ \frac{\pi}{2} - \theta_n -\frac{\sin(2 \theta_n)}{2} \right. \nonumber \\
&-& \left. \frac{C_2}{8 N C_3} \cot(\theta_n) \right].
\label{eq:2d_theta}
\end{eqnarray}

\noindent It is important to note that Eq.~(\ref{eq:2d_theta}) captures both the leading contribution found in Eq.~(\ref{eq:2D_leading_order}) and next order corrections in $\frac{\lambda_e}{\lambda_M}$. From Eq.~(\ref{eq:2d_theta}), it is easy to see that:

\begin{equation} 
n_{max} \sim N^2.
\end{equation}

In the limit $N^2 \gg n \gg N^{1/2}$, $\theta_n \approx \frac{\pi}{2} - \theta_0$, and one can show that:

\begin{eqnarray}
E_n = - \frac{(C_2^2 3 \pi)^{2/3}}{2^{5/3} a^2 C_1^{1/3} C_3^{2/3}} \left( \frac{n + \frac{1}{2}}{N^{1/2}} \right)^{2/3} e^{2h_N} \times \nonumber \\
\exp \left[ -2^{4/3} \frac{C_3^{1/3} (3 \pi)^{2/3}}{C_1^{1/3} C_2^{1/3}} \left( \frac{n + \frac{1}{2}}{N^{1/2}} \right)^{2/3} \right]. \
\end{eqnarray}

Finally, one can obtain an estimate for the semiclassical ground state energy using Eq.~(\ref{eq:2d_theta}). In order to do so, we note that when the quantum number $n$ is set to zero, and the zero point fluctuations are ignored, the right hand side of Eq.~(\ref{eq:2d_theta}) determines the semiclassical ground state energy. However, in this calculation, Eq.~(\ref{eq:2d_theta}) applies to the case when $ n \gg N^{1/2}$; the opposite limit of the semiclassical ground state. Therefore, we expect the correct N dependency to be captured by this estimate, but not the exact coefficient. The result of this analysis is:

\begin{equation}
E_0 = -\frac{3 C_2^2}{32 C_3 a^2}e^{2h_N}, \nonumber
\end{equation}

\noindent which is, as expected, the same as the semiclassical piece of Eq.~(\ref{eq:low_loying_spectrum_2d}), up to an overall constant.

\section{Lifetime of Excitations}

Up to this point, we have been focusing on the semiclassical solutions to the Hamiltonian, Eq.~(\ref{eq:hamiltonian}). However, as was seen earlier, the quantum correction to the Hamiltonian has a small imaginary piece. This imaginary correction implies that the semiclassical solutions are not stationary states, but have a finite lifetime. To obtain this lifetime, it is still possible to use the WKB method, except now there is a small imaginary piece to the potential $V = V(\lambda) + i \ Im \delta H$, and consequently, to the energy eigenvalue, $E = E_n + i \delta E$. The WKB formalism then states:

\begin{equation}
\left( n + \frac{1}{2} \right) \pi = \int_{\lambda_m}^{\lambda_M} \sqrt{2 \left(E_n + i \delta E - V(\lambda) - i \ Im \delta H \right)}.
\end{equation}

\noindent For very small values of $Im \delta H$, it is possible to Taylor series the above expression. Equating the real parts will yield the original quantization condition that has been utilized in the previous appendices. However, equating the imaginary parts will give:

\begin{equation}
\delta E = \frac{1}{\tau \left( E_n \right)} = \frac{\int d\lambda Im (\delta H) \left[2 (E_n - V(\lambda) )\right]^{-1/2}}{\int d\lambda \left[2 (E_n - V(\lambda)) \right]^{-1/2}}.
\label{eq:lifetime}
\end{equation}

\noindent Eq.~(\ref{eq:lifetime}) is identical to a semiclassical average of $Im (\delta H)$,  over a state with eigenvalue $E_n$. One can then substitute the potential from Eq~(\ref{eq:WKB}) into Eq.~(\ref{eq:lifetime}). The denominator of Eq.~(\ref{eq:lifetime}) then yields:

\begin{equation}
I = \int d\lambda \left[ \frac{C_2 N}{2 \lambda_M^2} - \frac{C_3 \epsilon N^2}{\lambda_M^d a^{\epsilon}} - \frac{C_2 N}{2 \lambda^2} + \frac{C_3 \epsilon N^2}{\lambda^d a^{\epsilon}} \right]^{-1/2},
\end{equation}

\noindent which can be arranged to give:

\begin{eqnarray}
I = \sqrt{\frac{\lambda_e^{\epsilon}(2 - \epsilon)\lambda_M^{4-\epsilon}}{2 C_2}} \int_0^1 dx \left[ 1 - x^{2 - \epsilon} \right. \nonumber \\
- \left. \left( \frac{\lambda_e}{\lambda_M} \right)^{\epsilon} \frac{1}{\lambda^{\epsilon}} (1-x^2) \right]^{-1/2} x^{1- \epsilon / 2},
\end{eqnarray}

\noindent where $\lambda_e$ is the semiclassical minimum of the potential, and $\lambda_M$ is the classical turning point of the potential. $\frac{\lambda_e}{\lambda_M}$ is still a small quantity, and, as was shown in Appendix C, generates a negligible contribution to the integral. Therefore, it is possible to keep only the zeroth order term in $\left( \frac{\lambda_e}{\lambda_M} \right)$, which is:

\begin{equation}
I = \sqrt{\frac{\lambda_e^{\epsilon}(2 - \epsilon)\lambda_M^{4-\epsilon}}{2 C_2}} \int_0^1 dx \frac{x^{1 - \epsilon/2}}{\sqrt{1-x^{2 - \epsilon}}}.
\label{eq:denominator}
\end{equation}

\noindent For the numerator, we note Eq.~(\ref{eq:imaginary_component}) depends on $\lambda^{d^2/2}$ and thus only adds an extra factor of $x^{d^2/2}$ to the integrand. Combining Eqs~(\ref{eq:denominator}), (\ref{eq:imaginary_component}), and (\ref{eq:lifetime}) gives the final expression for the lifetime:

\begin{eqnarray}
\tau^{-1}(E_n) &=& \frac{C_4}{2 a^2 C_3^{1-\epsilon/2}}N^{\epsilon / 2} (E a^2 )^{1 - \epsilon / 2} f_2 (\epsilon), \nonumber \\
f_2 (\epsilon) &=& \epsilon Im Fa \frac{\int_0^1 dx x^{(\epsilon - 1)(1- \epsilon / 2)}(1-x^{2- \epsilon})^{-1/2}}{\int_0^1 dx x^{1-\epsilon / 2} (1 - x^{2 - \epsilon} )^{-1/2}}, \nonumber \\
Im F_a &=& \frac{2 \pi^d}{\Gamma \left( \frac{d}{2} \right)} \int_0^{\sqrt{2}} dx x^{d-1} \sqrt{2x^2- x^4}.
\label{eq:lifetime_scaling}
\end{eqnarray}

\noindent $F_a$, defined in Eq.~(\ref{eq:lifetime_scaling}), is a smooth function of $\epsilon$, for all values of $\epsilon \geq 0$. After evaluating the above equations, one finds that the lifetimes for 1D and 2D are:

\begin{eqnarray}
\tau_{1D} (E_n) &=& \frac{3 \pi^{1/2} C_3^{1/2} a}{C_4 \ 2^{7/2}} \frac{1}{\sqrt{|E_n| N}}, \\
\tau_{2D} (E_n) &=& \frac{6 C_3}{\pi^3 C_4} \frac{1}{|E_n|}.
\label{eq:lifetimes_explicit}
\end{eqnarray}

\noindent An equivalent form for the lifetimes can be obtained by using the results Eqs.~(\ref{eq:high_lying_spectrum}) and (\ref{eq:high_lying_spectrum_2d}). In this case the lifetimes simplify to:

\begin{eqnarray}
\tau_{1D} (E_n) &=& \frac{3 \pi^{1/2} a^2}{8 C_4} \frac{1}{\sqrt{C_1 C_3}} \frac{\left( n + \frac{1}{2} \right) }{N^3}, \label{eq:lifetime_hl_1d} \\
\tau_{2D} (E_n) &=& \frac{12 C_3}{\pi^3 C_2 C_4} \frac{a^2}{N} e^{-2 h_N \sin^2 \theta_n} \tan^2(\theta_n), \label{eq:lifetime_hl_2d} \nonumber \\
\end{eqnarray} 

\noindent where $\theta_n$ in the 2d lifetime satisfies Eq.~(\ref{eq:2d_theta}).

These lifetimes can be understood from the semiclassical picture of the system and the dynamics of the phonons. The droplet size is of the same order of the scaling parameter, $\lambda$, which spends most of its time near the turning points of the potential. In the case of high-lying excitations, the system will spend most of its time near the turning point $\lambda_M$, which is assumed to be much larger than $\lambda_e$. At this scale, the phonons have a typical speed $v = \sqrt{|\mu|} \sim \sqrt{\frac{N}{\lambda_M^d}}$. At this scale, the speed of the dynamics slows down tremendously since the phonons are appreciably slower. Moreover, we note that larger energies correspond to larger values of $\lambda_M$, and one then expects that the lifetimes become longer for bound states of larger energy. 

For low-lying excitations, a similar calculation starting from Eq.~(\ref{eq:lifetime}) can be performed, but will not yield accurate results. For the ground state and low-lying excitations, the characteristic length scale of the system is equivalent to the length scale where the phonons become fully stable. As a result, the number of phonon modes that are unstable is small, and the characteristic length scale of these unstable phonons is that of the system size $\lambda$. As mentioned previously, fluctuations in this region are strongly dependent on the form of the scaling function used. As a result, applying our analysis to the lifetime of low-lying excitations is inadequate, as it relies heavily on the form of the scaling function.


\begin{thebibliography}{References}
\bibitem{diagrammatic}

The effective interaction potential for such a field defined at scale $L$ can also be computed using the same method that was employed to study unitary Bose gases in high dimensions~\cite{Borzov12,Mashayekhi13,Jiang13, Coleman73}. Indeed, the effective potential at scale $L=\lambda$ can be cast into the following form,





\begin{eqnarray}
V(L)& = &\frac{N}{L^2} N \tilde{g}(L) +V_{qc}(L), \nonumber\\
V_{qc}(L) &=&\frac{N}{L^2} \sum_{l \geq 1} \tilde{g}^l(L) \sum_{M\geq 3} C_l(M) (N \tilde{g}(L))^{M-1}. \nonumber
\end{eqnarray}

\noindent Here the first term in $V(L)$ is the Hartree-Fock energy. And $V_{qc}(L)$ represents the quantum corrections from $M$-body scattering processes with $l=1,2,3...$ internal loops, or virtual processes. $C_L(M)$ is a coefficient to be calculated.Since $\tilde{g}(L)$ vanishes as $L$ is much smaller than $a$, the contributions from $l$-loops are suppressed by a small factor $\tilde{g}^l(L)$. At small distance, we can safely neglect contributions with more than two loops and focus on the one-loop contributions. The evaluation of the one loop effects can be done straightforwardly and the result is fully consistent with Eq.(\ref{qc}).
\end{thebibliography}
\end{document}